\documentclass[12pt]{article}
\usepackage[normalem]{ulem}
\usepackage{amsfonts,slashed}
\usepackage{arydshln}
\usepackage[greek,english]{babel}
\usepackage{cancel}
\usepackage{upgreek}
\usepackage{float}
\usepackage{url}
\usepackage{latexsym}
\usepackage{mathtools}
\usepackage{amsmath, amscd}
\usepackage{amssymb}
\usepackage{epsfig}
\usepackage{latexsym,amssymb}
\usepackage{amsmath,amssymb,amsthm}
\usepackage{hyperref}
\usepackage{fontenc}
\usepackage{graphicx,tikz,shortvrb}
\usepackage{multirow}
\usepackage{xfrac}
\usepackage[T1]{fontenc}
\usepackage{textcomp}
\usepackage{lmodern}
\usepackage{stmaryrd}
\usepackage{cite}
\usepackage{mathrsfs,mathabx}
\usepackage[margin=20pt,small]{caption}
\usepackage{subcaption}
\usepackage{bbm}
\usepackage{accents}
\usepackage{geometry}
\usepackage{newlfont}
\usepackage{tabularx}
\usepackage{booktabs} 
\usepackage{witharrows}

\newcommand{\bea}{\begin{eqnarray}}
\newcommand{\eea}{\end{eqnarray}}
\newcommand{\ra}[1]{\renewcommand{\arraystretch}{#1}}

\begin{document}

\begin{titlepage}

\begin{center}
	{\LARGE \bf
Supersymmetric Null and Timelike Warped AdS and Strings  in \\ D=3, N=2 Gauged Supergravity  \\[1cm]}
	

	{ \bf 
	Nihat Sadik Deger\,$^{a,b,}{\!}$
		\footnote{\tt sadik.deger@bogazici.edu.tr},
	Ceren Ayse Deral\,$^{a,}{\!}$
		\footnote{\tt ceren.deral@bogazici.edu.tr}
	
		 \vskip .8cm}
	
	{\it ${}^a$ Department of Mathematics, Bogazici University, Bebek, 34342, Istanbul, T\"urkiye}\\[1.5ex] \ 
	{\it ${}^b$ Feza Gursey Center for Physics and Mathematics, Bogazici University, \\ Kandilli, 34684, Istanbul, T\"urkiye}\\[1.5ex] \ 
	 \\
	
\end{center}
\vfill

\begin{center}
	\textbf{Abstract}
	
\end{center}
\begin{quote}
We classify and construct supersymmetric solutions of  D=3, N=2 gauged supergravity extended with a Fayet-Iliopoulos term, and 
null and timelike warped AdS spacetimes are among them. From the first one, it is possible to obtain a well-defined black hole by periodic identification. We also find charged string solutions that interpolate between two supersymmetric AdS or Minkowski extrema of the scalar potential, one of which corresponds to a horizon with a singularity behind it, and the other determines the asymptotic geometry.


\end{quote}
\vfill
\setcounter{footnote}{0}

\end{titlepage}

\section{Introduction}
Three-dimensional (3D) gravity theories provide a fertile ground for investigating some difficult formal questions about quantum gravity, and they have important applications in AdS$_3$/CFT$_2$ holography. Over the years, several 3D supergravities have been constructed, and their properties and connections to other dimensions have been studied (for a review, see \cite{Sezgin:2023hkc}). However, still much remains to be done, in particular in classifying their supersymmetric solutions.

A systematic approach to this problem was provided by Tod \cite{Tod:1983pm, Tod:1995jf} where one uses Killing spinor bilinears to obtain algebraic and first-order differential conditions to be satisfied for such a solution to exist. Implementing these conditions simplifies the form of the physical fields, and most of the field equations are automatically fulfilled. Using this method in 3D off-shell supergravities led to some interesting backgrounds, including timelike, spacelike, and null warped AdS spacetimes \cite{Deger:2013yla, Deger:2016vrn, Alkac:2015lma, Deger:2018kur}.  For gauged 3D supergravities \cite{deWit:2003ja, deWit:2003fgi}, it was first used in \cite{Deger:2010rb, Deger:2015tra} for an N=8 gauged supergravity, but explicit solutions were found only in the ungauged limit. Recently, supersymmetric solutions of an $N=4$, gauged supergravity were classified and constructed using this technique \cite{Deger:2024xnd, Deger:2024obg}.
 
In this paper, we will apply this method to a 3D, $N=2$, matter coupled, U(1) gauged supergravity. This theory was constructed in \cite{Deger:1999st} and later extended in \cite{Abou-Zeid:2001inc} by including the Fayet-Iliopulos (FI) term \cite{Fayet:1974jb, Bagger:1982ab}. This brings two extra free parameters to its superpotential, and consequently, there are additional extrema. Supersymmetric backgrounds of \cite{Deger:1999st} were studied in \cite{Deger:2004mw, Deger:2006uc} using other approaches, whereas only the vortex solutions of  \cite{Abou-Zeid:2001inc} are known, which were found numerically in the same paper. $N=2$ supersymmetry requires the sigma model target space to be a K\"ahler manifold. We will focus on the model of \cite{Abou-Zeid:2001inc} with two real scalars and consider maximally symmetric $\mathbb{S}^2, \mathbb{H}^2$ and $\mathbb{R}^2$ target space geometries. As it is customary in this method, we classify our solutions
with respect to the norm of the Killing vector that is constructed from the Killing spinor as null or timelike. In each class, we find a large number of exact solutions which are listed in Table \ref{table}. Those for the $\mathbb{R}^2$ model are all new. For  $\mathbb{S}^2$ and  $\mathbb{H}^2$ models, in addition to generalizations of those obtained in  \cite{Deger:2004mw, Deger:2006uc} with FI terms, we find some novel solutions that are now allowed thanks to the extra parameters, such as timelike and null z-warped AdS which attracted a lot of attention in recent years and have found interesting applications, see \cite{Bieliavsky:2024vry} for a review. It is possible to obtain black holes from them by periodic identification \cite{Anninos:2008fx} similar to the well-known construction of the BTZ black hole from AdS$_3$ \cite{Banados:1992wn}. We show that by restricting parameters of the theory appropriately, our null z-warped solution can describe a physically well-defined black hole \cite{Anninos:2010pm}.

\begin{table*}[ht]
\centering
\ra{1.4}
\begin{tabular}{ m{0.35\textwidth} m{0.07\textwidth} m{0.15\textwidth} m{0.15\textwidth} m{0.12\textwidth} } \toprule
Solution & Eqn. & $\rho$ & $\theta$ & $A_\mu$\\
 \midrule
pp-wave in Minkowski or $AdS$  &\eqref{0betasoln}& $0$ & free & trivial \\
electromagnetic wave in Minkowski or $AdS$  &\eqref{constbetasoln}& $const.\neq0$ & $const.$ & nontrivial \\
null z-warped $AdS_3$  & \eqref{nwAdS}& $const.\neq0$ & $const.$ & nontrivial \\
string solution with waves  &\eqref{Gmetric2}& nontrivial & nontrivial & nontrivial \\
Euclidean Rindler spacetime  &\eqref{ERspace}& $0$ & free & trivial \\
timelike warped flat  &\eqref{twflat}& $const.\neq0$ & $0$ & nontrivial \\
timelike stretched $AdS_3$  & \eqref{twAdS}& $const.\neq0$ & $0$ & nontrivial \\
string-like solution  &\eqref{metrictimelike}& nontrivial & nontrivial & nontrivial \\
a timelike solution  &\eqref{atimelikesoln}& nontrivial & nontrivial & nontrivial \\
\bottomrule
\end{tabular}
\caption{The list of our supersymmetric solutions. They exist for all $\mathbb{R}^2, \mathbb{S}^2$ and $\mathbb{H}^2$ sigma models.
Our model has two scalars  $\{\rho, \theta\}$ and a vector field $A_\mu$.} \label{table}
\end{table*}

Unlike off-shell supergravity theories, where scalar fields are auxiliary, and hence their Lagrangian contains only a cosmological constant, gauged 3D supergravities have active scalars and typically include a scalar potential in their action.  For these theories, several supersymmetric string solutions supported by scalar fields have been found in the past; see, e.g., \cite{Deger:1999st, Berg:2001ty, Deger:2004mw, OColgain:2010yxb, Deger:2019jtl, Deger:2024xnd}. One common feature of them is that they all have a curvature singularity, some naked and some behind a horizon. For our model, we obtain string solutions with nontrivial gauge fields and, by making suitable coordinate transformations, show that they interpolate between two supersymmetric AdS or Minkowski extrema of the scalar potential.   
One of them corresponds to the horizon, and the other determines the asymptotic geometry. For the $\mathbb{H}^2$ and $\mathbb{R}^2$ sigma models, horizon exists only when the FI term is present.

The organization of our paper is as follows. We start with an introduction to the model \cite{Abou-Zeid:2001inc}  that we will work on and explain its connection to two other $N=2$ supergravity theories that were constructed earlier \cite{Izquierdo:1994jz, Deger:1999st}. In section \ref{sec3}, we make a general Killing spinor analysis using Tod's method \cite{Tod:1983pm, Tod:1995jf} and
show that one of the vectors that is constructed using the Killing spinors has to be null or timelike. We investigate the first option in section \ref{sec4}; and, in addition to wave-type solutions, we find null z-warped AdS and string solutions. We also study string backgrounds in some detail by introducing appropriate coordinate systems to look at their singularities and horizons. The timelike Killing vector case is worked out in section \ref{sec5}. We show that there are seven first-order differential equations to solve. We first look for exact solutions when one of the scalar fields is constant and find timelike warped AdS among them. We then consider solutions with circular symmetry with respect to one of the space coordinates and find "string-like" solutions whose metrics are almost identical with string solutions of section \ref{sec4} but have one extra piece that has no effect on their curvature invariants. We conclude with some future directions and comments in section \ref{sec6}.

\section{The Model}

We will study supersymmetric solutions of the U(1) gauged  D=3, $N=2$ supergravity constructed in \cite{Abou-Zeid:2001inc}, which is a generalization of the model of \cite{Deger:1999st} with Fayet-Iliopoulos (FI) term  \cite{Fayet:1974jb, Bagger:1982ab}. Bosonic fields of the model are $n$ complex scalars, a gauge field $A_\mu$,  and the dreibein. Dynamics of the scalar fields are described by a nonlinear sigma model whose target space has to be a $2n$-dimensional K\"ahler manifold for the $N=2$ supersymmetry. We will take $n=1$ and denote the single complex scalar as $\phi = R e^{i\theta}$. Assuming that the K\"ahler potential $K$ of the target space depends only on $|\phi|=R$, the bosonic Lagrangian of the model takes the form
\begin{align}\label{Lagrangian1}
\mathcal{L}= \sqrt{-g} \left( \frac{1}{4}\mathcal{R} -G\,(\partial_\mu R \partial^\mu R + R^2 D_\mu\theta D^\mu\theta)  - \frac{g_0}{8} \epsilon^{\mu\nu\rho} A_\mu F_{\nu\rho} + g_0^2\, V \right)  \, ,
\end{align}
where $G(R)$ is the metric of the sigma model target space and
\begin{align}
    D_\mu\theta=\partial_\mu\theta+g_0A_\mu\, .
\end{align}
Here, $g_0$ is the gauge coupling constant, and the ungauged model can be obtained by setting it to zero. The scalar potential $V(R)$ is given in terms of the superpotential $W(R)$ as:
\begin{align}
&V= 2W^2- \frac{G^{-1}}{4} (\partial_R W)^2\,. \label{v1}
\end{align}
Both the metric  $G(R)$ and the superpotential $W(R)$ can be expresssed in terms of a function $\mathcal{C}(R)$ as
\bea
G &=& -\frac{1}{2R} \partial_R \mathcal{C} \, , \\
W &=&\mathcal{C}^2 + b \, , \label{superpotential}
\eea 
where $b$ is an arbitrary real constant. The function $\mathcal{C}$ is related to the K\"ahler potential $K(R)$ of the target space as
\begin{align}
\mathcal{C} = -\frac{1}{2}R\partial_R{K}+c \, ,
\end{align}
where $c$ is a constant. In this paper, we will focus on maximally symmetric target spaces for the sigma model. This corresponds to choosing the K\"ahler potential as
\begin{align}
K(R) = \begin{cases} 
      \frac{\varepsilon}{a^2} \ln (1+\varepsilon R^2)\,,\, \text{for} \,\,  \mathbb{S}^2 \, (\varepsilon=1) \,\, \text{and}\,\, \mathbb{H}^2  \, (\varepsilon=-1) \,  , \\
     R^2\,,\, \text{for}  \,\,  \mathbb{R}^2 \, ,
   \end{cases} 
\end{align}
where $a^2$ is a constant and the parameter $\varepsilon=\pm 1$ labels $\mathbb{S}^2$ and $\mathbb{H}^2$ 
manifolds.
To bring the Lagrangian \eqref{Lagrangian1} into a more standard form, we make the change of variable
\begin{align}
R = \begin{cases} 
      \tanh (a\rho)\, , & \text{for } \mathbb{H}^2\,, \\
      \tan (a\rho), \text{ for } \rho\in (-\frac{\pi}{2a},\frac{\pi}{2a}), & \text{for } \mathbb{S}^2\,, \\
     \rho\,, & \text{for } \mathbb{R}^2\,,
   \end{cases} \label{Rrhotransf}
\end{align}
after which \eqref{Lagrangian1} becomes
\begin{align}
\mathcal{L}=\sqrt{-g} \left( \frac{1}{4} \mathcal{R} - \partial_\mu \rho \, \partial^\mu \rho - \frac{1}{4a^2} \mathcal{S}^2\,D_\mu\theta D^\mu\theta -\frac{g_0}{8} \, \epsilon^{\mu\nu\rho} A_\mu F_{\nu\rho} +g_0^2\, V(\rho) \right)   \,, \label{Lagrangian2}
\end{align}
where
\begin{align}
\mathcal{S} = \begin{cases} 
      \sinh (2a\rho)\,, & \text{for } \mathbb{H}^2\,, \\
      \sin (2a\rho)\,, & \text{for } \mathbb{S}^2\,, \\
     2a\rho\,, & \text{for } \mathbb{R}^2\, ,
   \end{cases} \label{defnS}
\end{align}
and we take $a>0$ without loss of generality. The scalar potential \eqref{v1} is now
\begin{align} \label{potential}
V=2(\mathcal{C}^2+b)^2 - \frac{1}{a^2}\mathcal{S}^2\mathcal{C}^2\,, 
\end{align}
where
\begin{align}
\mathcal{C}  = \begin{cases} 
      - \frac{1}{2a^2}\cosh (2a\rho)+ x_0\,=\, -\frac{1}{a^2}\sinh^2(a\rho)+c\,, & \text{for } \mathbb{H}^2\,, \\
     \frac{1}{2a^2} \cos (2a\rho)+ x_0 \,=\, -\frac{1}{a^2}\sin^2(a\rho)+c\,, & \text{for } \mathbb{S}^2\,, \\
     -\rho^2 +c \,, & \text{for } \mathbb{R}^2\,.
   \end{cases} \label{defnC}
\end{align}
Here we introduced a new constant $x_0$  for the $\mathbb{S}^2$ and $\mathbb{H}^2$ manifolds  as
\begin{align}
x_0=c-\frac{\varepsilon}{2a^2} \, . \label{x0}
\end{align}
Interestingly, for the following choice of parameters $\{\varepsilon=-1, a^2=-b=1/2, x_0=0\}$ the potential \eqref{potential} takes the constant value $V=1/2$,
although the superpotential $W=\cosh(4a\rho)/2$ remains nontrivial.
Let us also note some useful identities,
\begin{align}
& \varepsilon \mathcal{S}^2 = 1-4a^4(\mathcal{C}-x_0)^2\,, \hskip2em (\text{for }\mathbb{H}^2 \text{ and }\mathbb{S}^2) \notag\\
& \partial_\rho \mathcal{S} = 4\varepsilon a^3(\mathcal{C}-x_0)\,, \hskip3.6em (\text{for }\mathbb{H}^2 \text{ and }\mathbb{S}^2) \notag\\
&\partial_\rho \left(\frac{\mathcal{S}}{\mathcal{C}-x_0}\right)= \frac{\varepsilon}{a \,(\mathcal{C}-x_0)^{2}}   \,, \hskip0.8em (\text{for }\mathbb{H}^2 \text{ and }\mathbb{S}^2)  \notag \\
& \partial_\rho \mathcal{C} = -\frac{1}{a}\mathcal{S} 
   \, . \label{identities}  
\end{align}

\begin{figure}[ht]
\begin{minipage}{0.32\linewidth}
\includegraphics[width=\textwidth]{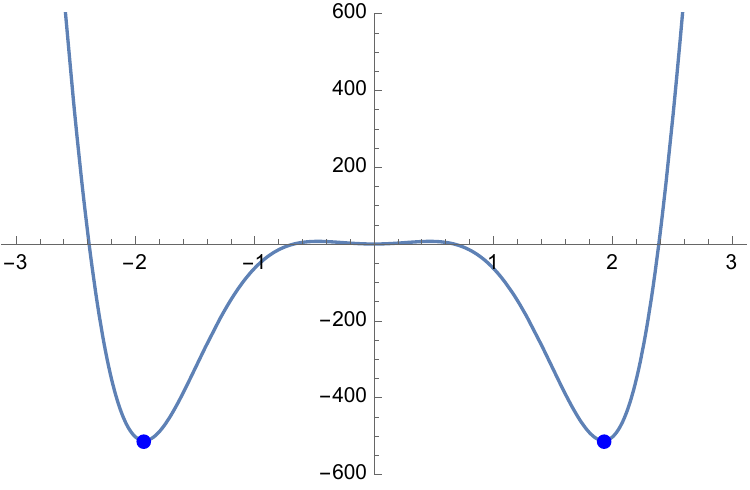}
\caption*{$c^2+b=0$ with $b\neq0$}
\label{fig:figure1}
\end{minipage}%
\hfill
\begin{minipage}{0.32\linewidth}
\includegraphics[width=\textwidth]{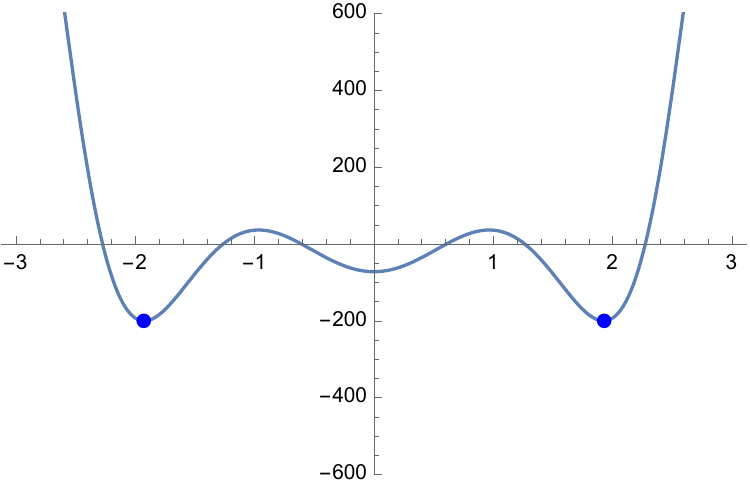}
\caption*{$c^2+b\neq0$ with $b\neq0$}
\label{fig:figure2}
\end{minipage}%
\hfill
\begin{minipage}{0.32\linewidth}
\includegraphics[width=\textwidth]{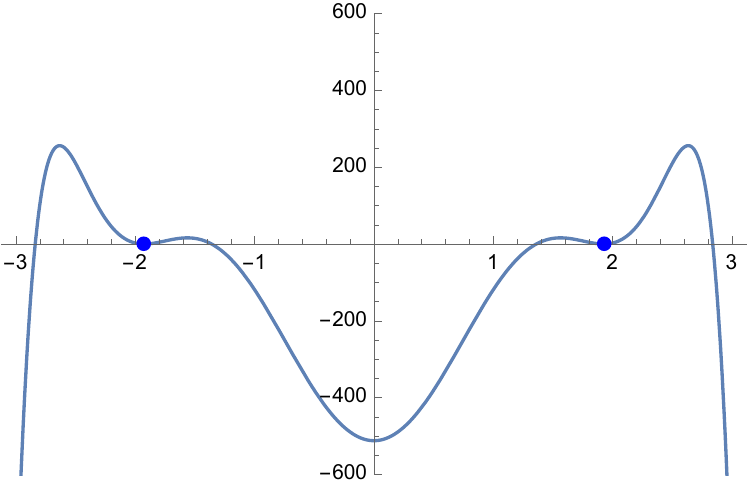}
\caption*{$c^2+b\neq0$ with $b=0$}
\label{fig:figure3}
\end{minipage}
\caption{Graph of the $-V(\rho)$ given in \eqref{potential} for the $\mathbb{H}^2$ sigma model with $c>0$. There is always a
supersymmetric vacuum at $\rho=0$ and a pair of supersymmetric extrema located at $\mathcal{C}=0$ (indicated with blue dots) when $c>0$.
They are Minkowski or AdS, depending on whether ($c^2+b$) and $b$
are vanishing or not, respectively. These characteristics are the same for the $\mathbb{S}^2$ and $\mathbb{R}^2$ sigma models.} \label{fig}
\end{figure}

The supersymmetric extrema of the potential \eqref{potential} are at $\partial_\rho W= -2\mathcal{C}\mathcal{S}/a=0$.  From \eqref{defnS} and \eqref{defnC} we find that they are located at
\begin{align}\label{vacua}
& i)\,\,\, \rho =0: \, c^2\neq -b \implies \text{AdS} \,, \, c^2=-b \implies \text{Minkowski} \, ,  \notag \\
& ii)\, \mathcal{C}=0 \, ,c>0: \,  b \neq 0 \implies \text{AdS} \, , \,  b=0 \implies  \text{Minkowski} \, .
\end{align}
So, there is always a fully supersymmetric AdS or Minkowski vacuum at $\rho=0$, but the other supersymmetric vacua exist only when $c>0$. Such vacua come as a pair since the potential \eqref{potential} is invariant under $\rho \rightarrow -\rho$. There can be only one pair, which can be seen from \eqref{defnC}, and such a pair always exists when $c>0$. They are located at $\rho= \pm \sqrt c$ for the $\mathbb{R}^2$ model. Meanwhile, for  the $\mathbb{H}^2$, we need $x_0>1/(2a^2)$   from \eqref{x0}. For  $\mathbb{S}^2$, since $\rho\in (-\frac{\pi}{2a},\frac{\pi}{2a})$ the function  $\cos(2a\rho)$ is either positive or negative.  If we assume $\cos(2a\rho) >0$
then we need  $-1/(2a^2)<x_0 <0$, otherwise  $ 0< x_0<1/(2a^2)$. Finally, note that when the vacuum at $\rho=0$ is Minkowski, then the others should be AdS; but when the vacuum at $\rho=0$ is AdS,  the others can be AdS or Minkowski, when they exist. We illustrate these in Figure \eqref{fig}.

The $\mathbb{S}^2$ and $\mathbb{H}^2$ models have 3, i.e. \{$c$ or $x_0$, $b$, $a^2$\}, and the $\mathbb{R}^2$ model has 2, namely \{$c$, $b$\}, arbitrary constants.  The constant $c$ comes from the U(1) isometry of the K\"ahler potential and  
gives rise to the so-called Fayet-Iliopolous terms \cite{Bagger:1982ab}, and the constant $b$ is due to the freedom in defining the superpotential \eqref{superpotential}. Finally, $\kappa=4\varepsilon a^2$ is the Gaussian curvature of the $\mathbb{S}^2 \, (\varepsilon=1)$ and $\mathbb{H}^2\, (\varepsilon=-1)$ manifolds.  Since $\mathbb{S}^2$ is compact, it can not be covered by a single coordinate patch, and  the global consistency of the model requires quantization of the gravitational coupling constant \cite{Witten:1982hu}
which in our case gives the condition  $1/a^2=integer$ \cite{Deger:1999st}.  

From \eqref{Lagrangian2}, one finds the field equations of the model as
\begin{align} \label{fieldeq}
&\nabla_\mu \nabla^\mu \rho=  \frac{1}{4a^2}\mathcal{S} \, \partial_\rho \mathcal{S} \, D_\mu \theta D^\mu\theta - \frac{g_0^2}{2} \, \partial_\rho V  \,, \notag\\
& \nabla_\mu \big(\mathcal{S}^2  \, D^\mu\theta \big)=0 \,, \notag\\
& \epsilon^{\mu\nu\rho} F_{\nu\rho} = -\frac{2}{a^2}\mathcal{S}^2 \, D^\mu\theta  \,, \notag\\
&\mathcal{R}_{\mu\nu} = 4 \partial_\mu \rho \, \partial_\nu \rho + \frac{1}{a^2} \mathcal{S}^2 \,D_\mu\theta D_\nu\theta  - \, 4g_0^2 \, V g_{\mu\nu}  \,.
\end{align}
Note that setting $A_\mu=\theta=0$ is a consistent truncation. 

Finally, for finding supersymmetric solutions of the model, we set the fermions, that is, gravitino $\psi$  and matter fermion $\lambda$, and their supersymmetry transformations to zero:
\begin{align}
&\delta \psi= \nabla_\mu \epsilon + i\{ (\mathcal{C}-c) D_\mu\theta + \,g_0 \,c\,A_\mu \} \epsilon -g_0W\gamma_\mu \epsilon=0 \,, \notag\\
&\delta \lambda = (\partial_\mu \rho + \frac{i}{2a}\, \mathcal{S}D_\mu \theta) \gamma^\mu \epsilon - \frac{g_0}{a} \mathcal{C}\mathcal{S} \epsilon=0\,, \label{bpseqnsroot}
\end{align}
where $\epsilon$ is a complex spinor and $\nabla_\mu \epsilon=(\partial_\mu+ \frac{1}{4}\,\omega_\mu^{\,\,\,\, ab}\, \gamma_{ab}) \epsilon$.

It is easy to see that the $\mathbb{R}^2$ model above can be obtained by taking the $a \rightarrow 0$ limit in the  $\mathbb{S}^2$ or $\mathbb{H}^2$ models without scaling the fields. The $\mathbb{S}^2$ and $\mathbb{H}^2$ models with $x_0=b=0$ were constructed earlier in \cite{Deger:1999st} where  a flat sigma model
was obtained by taking the limit  $a \rightarrow 0$ but it was different than the above one \cite{Deger:1999st, Deger:2000as}. To see the reason
let us make constant scalings $A_\mu \to \frac{A_\mu}{k} \, , \, g_0 \to g_0 k^2 \, , \, b \to \frac{b}{k^2} \, , \,  c \to \frac{c}{k}$ while keeping scalar fields and fermions the same, and then send $k \to 0$ in the Lagrangian \eqref{Lagrangian2} which gives
\begin{align}
\tilde{\mathcal{L}}=\sqrt{-g} \left( \frac{1}{4} \mathcal{R} - \partial_\mu \rho \, \partial^\mu \rho - \frac{1}{4a^2} \mathcal{S}^2\, \partial_\mu\theta \partial^\mu\theta  -\frac{g_0}{8} \, \epsilon^{\mu\nu\rho} A_\mu F_{\nu\rho}  +2g_0^2\, (c^2 + b)^2 \right)  \, . \label{Lagrangian3}
\end{align}
Note that here $\theta$ is uncharged with respect to the R-symmetry group U(1) and as a result, the potential \eqref{potential} has reduced to a cosmological constant. This model was constructed earlier in \cite{Izquierdo:1994jz}. After this procedure, the target space geometry remains the same in general. However, if one starts with the  $\mathbb{S}^2$ or $\mathbb{H}^2$ model with $x_0=0$  so that $c=\varepsilon/2a^2$  is not an independent parameter but depends on $a^2$, then $a \to 0$ limit is equivalent to first scaling $a\to a\sqrt k $ and then sending $k\to 0$. Since this results in transforming  $ c \to \frac{c}{k}$,  one also has to do other scalings above for consistency, which gives the Lagrangian \eqref{Lagrangian3} with flat target space that was found in \cite{Deger:1999st, Deger:2000as}. This clarifies the connection between different 3D, $N=2$ supergravities that exist in the literature \cite{Izquierdo:1994jz, Deger:1999st, Abou-Zeid:2001inc}.
After this short introduction, we are now ready to search for supersymmetric solutions of this model \cite{Abou-Zeid:2001inc}.

\section{Killing Spinor Analysis} \label{sec3}
For our analysis of the Killing spinor equations \eqref{bpseqnsroot}, it is more convenient to work with Majorana spinors\footnote{We follow the same conventions with \cite{Deger:2024xnd, Deger:2024obg}. Three-dimensional tangent space indices range from 0 to 2. The Levi-Civita tensor is $\epsilon^{\mu\nu\sigma}$ and the Levi-Civita symbol is $\varepsilon^{012}=-1$ which are related as $\epsilon_{\mu\nu\sigma}=\sqrt{-g}\varepsilon_{\mu\nu\sigma}$, $\epsilon^{\mu\nu\sigma}=(\sqrt{-g})^{-1}\varepsilon^{\mu\nu\sigma}$.   Gamma matrices with tangent space indices are chosen as: 
$\gamma^0=i\sigma^2\,,\, \gamma^1 =\sigma^3\,,\,\gamma^2=\sigma^1$, where $\sigma$'s are the Pauli matrices. The charge conjugation matrix is $C=\gamma^0$. We have $\bar{\lambda}=\lambda^{\dagger} C$
and a Majorana spinor satisfies $\lambda^*=-i\lambda$.} $\lambda_A$ ($A=1, 2$) so that
\begin{align}
\epsilon \, = \, \lambda_1 \, + \, i \, \lambda_2 \,. \label{sp}
\end{align}
Then, the BPS conditions \eqref{bpseqnsroot} take the form
\begin{align}
& 0 \, = \, \nabla_\mu \lambda_A + X_\mu \epsilon_{AB} \lambda^B - g_0 W \gamma_\mu \lambda_A \,, \label{kse} \\
& 0 \, = \, \partial_\mu\rho \gamma^\mu \lambda_A + \frac{1}{2a} \mathcal{S} D_\mu\theta \gamma^\mu \epsilon_{AB} \lambda^B +  \frac{g_0}{2}  \frac{\partial W}{\partial \rho} \lambda_A \,, \label{bps} 
\end{align} 
where $\epsilon_{AB}=-\epsilon_{BA}$ and $\epsilon_{12}=\epsilon^{12}=-1$ and 
\begin{align} \label{X}
X_\mu=  (\mathcal{C}-c) D_\mu\theta + g_0 c A_\mu \, .
\end{align}
We will use Tod's Killing spinor bilinears method \cite{Tod:1983pm, Tod:1995jf} to classify and construct all supersymmetric solutions of this model.  For that purpose, we assume that there is a Killing spinor $\epsilon$ that solves these supersymmetry conditions \eqref{kse}-\eqref{bps} and  define real bilinears
\begin{align}
F^{AB}&=\bar{\lambda}^A\lambda^B=-F^{BA} \, ,  \\
V_\mu^{AB}&=\bar{\lambda}^A\gamma_\mu\lambda^B=V_\mu^{BA} \, .
\end{align}
Since indices $A, B, ...$ run from 1 to 2, the antisymmetry of $F^{AB}$ implies that
\begin{equation}
    F^{AB}=-f\epsilon^{AB} \, , 
\end{equation}
where $f$ is a function. This result and the structure of the BPS equations \eqref{kse}-\eqref{bps} are precisely the same with those studied in \cite{Deger:2024xnd, Deger:2024obg} for the D=3, $N=4$, U(1)$\times$ U(1) gauged supergravity with 
modified definitions of the superpotential $W$ \eqref{superpotential}  and $X_\mu$ \eqref{X}. Therefore, we will be brief on technical details in deriving some of our results below, which can be found in these papers. 

We first define vectors
\begin{align} \label{vectors}
V_\mu=V_\mu^{11}+V_\mu^{22} \,, \quad
K_\mu=V_\mu^{11}-V_\mu^{22} \,, \quad
L_\mu=2V_\mu^{12} \,,
\end{align}
which satisfy
\begin{align}
&V^\mu K_\mu = V^\mu L_\mu=K^\mu L_\mu=0 \notag \,, \quad 
V_{[\mu}K_{\nu]} = \epsilon_{\mu\nu\sigma}f L^\sigma \,, \\
&V^\mu V_\mu=-K^\mu K_\mu=-L^\mu L_\mu=-4f^2 \,.
\label{algebraic}
\end{align}
When $f\neq 0$, they form a 3-dimensional orthogonal basis for the Lorentzian spacetime, and when $f=0$, one can choose $V_\mu=K_\mu$ and $L_\mu=0$ without loss of generality \cite{Deger:2010rb}.  From \eqref{kse} one also gets
\begin{align} 
&\nabla_\mu V_\nu =-W\epsilon_{\mu\nu\sigma} V^\sigma \, , \label{vd} \\
&\nabla_\mu K_\nu =-W\epsilon_{\mu\nu\sigma} K^\sigma +2X_\mu L_\nu \, , \label{kd} \\
&\nabla_\mu L_\nu =-W\epsilon_{\mu\nu\sigma} L^\sigma -2X_\mu K_\nu \label{ld} \, .
\end{align}
The first one implies that $\nabla_{(\mu} V_{\nu)}=0$, that is, $V$ is either a timelike ($f\neq 0$) or a null ($f=0$) Killing vector. Moreover, by contracting it with $V^\nu$, we obtain
\begin{equation}
    \partial_\mu f=0 \, .
\end{equation}
Now, \eqref{bps} leads to
\begin{align}
& 0 \, = \, \partial_\mu \rho V^\mu \,, \label{Lie1}\\
& 0 \, = \, \epsilon^{\nu\mu\sigma} \partial_\mu \rho V_\sigma - \frac{f}{a} \mathcal{S}D^\nu\theta  - \frac{g_0}{a}  \mathcal{C}\mathcal{S} V^\nu \,, \label{bps2} \\
& 0 \, = \, 2 f \partial^\nu \rho + \frac{1}{2a}  \mathcal{S} \epsilon^{\nu\mu\sigma} D_\mu\theta V_\sigma \,. \label{bps3}
\end{align}
The first line simply means that the Lie derivative of the field $\rho$ along the Killing vector $V$ is zero, that is, $\mathcal{L}_V \rho=0$. Now, to have this property also for the remaining fields, we make the gauge choice 
\begin{align}
V^\mu A_\mu =  4f\mathcal{C} \,, \label{gauge}
\end{align}
which differs from the choice made in \cite{Deger:2024xnd, Deger:2024obg}. After this, we have
\begin{align}
\mathcal{L}_V \theta =\mathcal{L}_V A  =0\,, \label{Lie2}
\end{align}
where we used the vector field equation in \eqref{fieldeq} as well.

Solutions are, in general, half-supersymmetric, except for Minkowski and AdS spacetimes, which are fully supersymmetric. 
The analysis from this point onward depends on whether the constant $f$ is vanishing or not. We will now investigate these two cases separately.

\section{Null Killing Vector} \label{sec4}
In this part, we set $f=0$ and denote the null Killing vector as $V=\partial_v$.  Due to \eqref{Lie1} and \eqref{Lie2}, none of the physical fields depend on the $v$-coordinate. Now, assuming that the scalar field $\rho$ depends only on the radial $r$-coordinate , that is, $\rho=\rho(r)$, the most general spacetime metric with $V=\partial_v$ as its null Killing vector has the form \cite{Deger:2024xnd}
\begin{align}
ds^2=dr^2+2e^{2U(r)}dudv+e^{2\beta(u,r)}du^2 \, . \label{Gmetric}
\end{align}
Here, the function $\beta(u,r)$ is to be determined, and $U(r)$ is fixed by the superpotential $W$ \eqref{superpotential}   as
\begin{align}
\partial_r U(r)= 2g_0 W \, . \label{U}
\end{align}
The computation of the Killing spinors goes through exactly as in \cite{Deger:2024xnd}, and one finds
\begin{align}
    \lambda^A= (1+i)\, e^{U-\tfrac{1}{2}\beta}\, \lambda_0^A \, ,
\end{align}
where $\lambda_0^A$ is a constant real spinor that satisfies $(\gamma^1-\gamma^0)\lambda_0^A=0$. 

Now, the remaining BPS conditions that follow from \eqref{ld} ,\eqref{bps2} and \eqref{bps3} are
\begin{align}
& \mathcal{S}D_r\theta \label{Stheta0} = 0 \,, \\
& (\mathcal{C}-c) D_r\theta  = - g_0 c A_r\,, \label{X1}\\
& (\mathcal{C}-c) D_u\theta =  - g_0 c A_u \,, \label{X2} \\
& \partial_r \rho  \, =\, \frac{g_0}{a} \mathcal{C}\mathcal{S} \,. \label{rhoder0}
\end{align}
With these conditions, the field equations for $\rho$ and $\theta$ in \eqref{fieldeq} are identically satisfied.  Due to our gauge choice \eqref{gauge}, we have $A_v=0$ which sets $F_{vr}=F_{vu}=0$, and the only remaining vector field equation is
\begin{align}
& F_{ur} = - \frac{1}{a^2} \mathcal{S}^2 D_u\theta   \label{feFur}\,.
\end{align}
Finally, it can be shown that all Einstein's field equations hold except for $R_{uu}$, which gives the following condition for the metric function $\beta(u, r)$ 
\begin{align}
\partial_r \big( e^{2U} \partial_r (e^{2\beta-2U})\big) &= - \frac{2}{a^2}\mathcal{S}^2 (D_u\theta)^2\,. \label{Ruu0}
\end{align}
When the right-hand side of this equation is zero, setting $e^{2\beta}=0$ is a solution. Its homogeneous solutions can actually be found using the Garfinkle-Vachaspati solution-generating method \cite{Garfinkle:1990jq},
starting with the same metric with $e^{2\beta}=0$ (see \cite{Deger:2004mw, Deger:2024xnd} for details), and they are locally trivial \cite{Gibbons:2008vi}.  This method is applicable when the initial solution has a null Killing vector.

Now, \eqref{Stheta0} implies that either $S=0$ (which is equivalent to $\rho=0$) or $D_r\theta=0$. With the first option, the scalar field $\theta$ decouples from the model, and the vector field becomes pure gauge. The latter implies that either $c=0$ or $A_r=0$ from \eqref{X1}. When $c=0$ again the vector field is pure gauge and we have $D_u\theta=0$ from \eqref{X2} (here if $\mathcal{C}=0$ then $\rho=0$). This analysis shows that, without loss of generality, we can choose 
\begin{align}
A_r=0  \quad \, , \quad \theta=\theta(u) \, ,
\end{align}
after which \eqref{Stheta0} and \eqref{X1} are automatically satisfied.  

To summarize, for a supersymmetric solution in this class, we only have to solve 4 equations \eqref{X2}-\eqref{Ruu0} for the fields $\{\rho(r), A_u(r,u), \theta(u), \beta(u,r)\}$. The spacetime metric is given by \eqref{Gmetric} where the metric function $U(r)$ should satisfy \eqref{U}. We will now analyze these equations in 3 parts: i) $\rho=0$, ii) $\rho =constant\neq 0$, iii) otherwise.

\subsection{\texorpdfstring{$\rho\,=0$}{rho=0}}

In this case, from \eqref{superpotential} and \eqref{defnS}-\eqref{defnC} we have
\begin{align} \label{rho=0}
\mathcal{S}=0 \quad \,, \quad \mathcal{C}=c  \quad \,, \quad W=c^2+b \quad \,, \quad V=2(c^2+b)^2 \, .
\end{align}
Equations \eqref{X1} and \eqref{feFur} show that the vector field is pure gauge, which we choose as $A_u=0$, and the field $\theta$ remains free. Now, \eqref{U} is solved as 
\begin{align}
U(r)=2g_0(c^2+b)r  \,, \notag\\
\end{align}
where we set an arbitrary additive constant to zero by scaling the $v$-coordinate. Finally, solving \eqref{Ruu0} for $e^{2\beta}$ we find
\begin{align}
ds^2 = dr^2 + 2 e^{ 4g_0 (c^2+b) r} dudv +e^{2\beta} du^2 \label{Adspp} \, ,
\end{align} 
where
\begin{align}
e^{2\beta} = \begin{cases} 
      c_3(u)r + c_4(u)\,, & \text{if } c^2 + b=0\,, \\
      c_3(u) e^{4g_0 (c^2+b) r} + c_4(u) \,, & \text{if } c^2 + b\neq0\,.
   \end{cases} \label{0betasoln}
\end{align}
Its curvature scalar is $\mathcal{R} \, =  -24g_0^2(c^2+b)^2 $.
This is a pp-wave in Minkowski or AdS spacetime, depending on whether $(c^2 + b)$ vanishes or not, respectively.
Although $c_3(u)$ and $c_4(u)$ terms can be generated by the Garfinkle-Vachaspati method and locally trivial, they may have global significance. For instance, when  $c^2 +b \neq 0 $ and $c_3$ and $c_4$ are constant, the metric is of Kaigorodov type \cite{Cvetic:1998jf, Brecher:2000pa} which is equivalent to the extremal BTZ black hole \cite{Banados:1992wn} with mass $M_0=2g_0 (c^2+b)c_4$ and angular momentum $J=c_4$.

\subsection{\texorpdfstring{$\rho=constant\neq0$}{rho const}}
When $\rho$ is a nonzero constant we have $\mathcal{S}=constant\neq 0$ from \eqref{defnS}. Then \eqref{rhoder0} implies that $\mathcal{C}=0$, which means that these solutions emerge at the supersymmetric extrema of the potential \eqref{potential} that are not at the origin.
From \eqref{superpotential} and \eqref{defnS}-\eqref{defnC} we obtain
\begin{align}
\mathcal{C}=0 \quad \,, \quad W=b \quad \,, \quad V=2b^2  \quad \,, \quad c> 0 \, .
\end{align}
With these, \eqref{U}, \eqref{X2} and \eqref{feFur} are solved as
\begin{align}
U(r)=2g_0b r \quad , \quad \theta(u)=constant \quad , \quad A_u(u,r)=Q(u)e^{2\eta g_0r} \, ,
\end{align}
where $Q(u)$ is an arbitrary function, and we introduced the nonzero constant
\begin{align} \label{sigma}
\eta \,=\, \frac{1}{2a^2}\mathcal{S}^2 \,=\, \begin{cases} 
     \frac{\varepsilon}{2a^2} (1-4a^4x_0^2) \,, & \text{for } \mathbb{S}^2 \, \text{and } \mathbb{H}^2 \, , \\
     2c \,, & \text{for } \mathbb{R}^2\,.
   \end{cases}
\end{align}
Then, solving \eqref{Ruu0}, we find the following 2 types of solutions:

\

\textbf{\underline{Case 1: Electromagnetic Waves}}

\

 When $W=b=\eta \neq 0$ or $W=b=0$ the metric is
\begin{align}\label{LogAdSpp}
ds^2 = dr^2 + 2 e^{4g_0br} dudv + e^{2\beta}  du^2 \, ,
\end{align} 
where
\begin{align}
e^{2\beta} = \begin{cases} 
     c_3(u)r+ c_4(u) - \frac{Q^2(u)}{4\eta}  \; e^{4g_0\eta r} \,, & \text{if } W=b=0\,, \\
      c_3(u) e^{4g_0br} +c_4(u) -g_0Q^2(u) r e^{4g_0br} \,, & \text{if } W=b=\eta \neq 0 \,.
   \end{cases} \label{constbetasoln}
\end{align}
Its curvature scalar is $\mathcal{R}  =  - 24g_0^2 \, b^2$ . Here, the locally nontrivial piece is the $Q^2(u)$ term, and the solution represents an electromagnetic wave in AdS (when $W=b=\eta \neq 0$) or Minkowski (when $W=b=0$) spacetime. When the vector field is switched off, the form of the solution coincides with the previous one given in \eqref{Adspp}.

\

\textbf{\underline{Case 2: Null z-warped AdS$_3$}}

\

When $W=b\neq\eta$, $\eta\neq 0$ and $b\neq0$ we find
\begin{align}
ds^2 = dr^2 + 2 e^{4g_0br} dudv +  [c_3(u) e^{4g_0br} +c_4(u) + \frac{Q^2(u)}{4( b-\eta)}   e^{4g_0\eta r}  ]du^2 \, ,
\label{AdSpp2}
\end{align} 
with
\begin{align}
\mathcal{R} \, = \, - 24g_0^2 \, b^2 \,.
\end{align}
This solution generically corresponds to an AdS wave.  However, if we choose $c_3(u)$, $c_4(u)$ and $Q(u)$  as constants and perform the change of coordinates 
\begin{align}
y=e^{4g_0br}  \quad , \quad \phi= \frac{u}{2g_0b} \quad , \quad   t= \frac{v}{2g_0b}
\end{align}
we get
\begin{align} \label{nwAdS}
ds^2= \frac{1}{4g_0^2b^2} \left( \frac{dy^2}{4y^2} + 2y dtd\phi +\left[\beta y + \alpha^2 + \frac{Q^2}{4(b-\eta)}y^{\eta/b}\right] d\phi^2\right) \, ,  
\end{align}
where $\alpha=c_3$, $\beta=c_4$ and Q are constants. The gauge field takes the form
\begin{align}
A= \frac{Q}{2g_0b} y^{\eta/(2b)} d\phi \, . \label{gf}
\end{align}
When $\alpha=\beta=0$ and $Q\neq 0$, \eqref{nwAdS} is called the {\it null z-warped AdS} spacetime with
\begin{align}
z=\frac{\eta}{b} \, .
\end{align}
This choice is special since then both the metric and the gauge field are invariant under the following anisotropic scaling
\begin{align} \label{ansc}
y \rightarrow  y \, \lambda^2 \quad , \quad \phi \rightarrow \phi\, \lambda^{-z} \quad , \quad t \rightarrow t \, \lambda^{z-2}   \, .
\end{align}
Depending on the sign of the $y^z$ term in \eqref{nwAdS}, this solution is labeled with "plus" or "minus" respectively. If we also fix the scaling exponent to $z=2$ so that the $t$-coordinate does not scale, then this solution is called {\it null warped AdS or Schr\"odinger spacetime} \cite{Duval:2024eod}.  

As was shown in \cite{Anninos:2010pm}, the metric \eqref{AdSpp2} describes a physically well-defined black hole if the $\phi$ coordinate is periodic ($\phi \sim \phi + 2\pi$) and
\begin{align}
z\geq 2 \quad , \quad \beta \geq 2|\alpha| \quad , \quad \beta\neq 0 \quad , \quad \alpha \neq 0 \quad , \quad Q^2= 4(b-\eta)  \, .
\end{align}
Note that after this identification, the scaling symmetry \eqref{ansc} is broken and the sign of the $y^z$ term in \eqref{nwAdS} has to be  "plus", which implies $b>\eta$. This requirement, together with the $z\geq 2$ condition, can be realized if we have the restriction $\eta \leq 2b < 0$ on the parameters of the model, where $\eta$ is defined in \eqref{sigma}. When this condition is satisfied, this model admits a null z-warped AdS black hole. 
 
Supersymmetric null z-warped AdS solutions appeared in off-shell supergravities before \cite{Deger:2013yla, Deger:2016vrn, Deger:2018kur}. However,  to the best of our knowledge, this is the first appearance of supersymmetric null z-warped AdS black hole in a gauged supergravity. Null warped AdS (i.e. $z=2$) is a solution of a D=3, $N=4$ gauged supergravity \cite{Deger:2024xnd} but with "minus" sign for the $y^z$ term in \eqref{nwAdS}, and hence it is not suitable for a black hole construction.  Such supersymmetric black holes \cite{Deger:2024dbz} also exist in Minimal Massive Supergravity \cite{Deger:2022gim, Deger:2023eah}. In that case, the warping is due to the higher curvature terms in the theory. Whereas, here it is created by the gauge field \eqref{gf}.

\subsection{String Solutions} \label{sectstr}
We now consider the $\rho = \rho(r) \neq constant$ case.  Here, instead of using the $r$-coordinate as the radial coordinate defined in \eqref{Gmetric}, it is possible to use the field $\rho$ (or the functions $\mathcal{C}$ or $\mathcal{S}$) as the radial coordinate via the BPS equation \eqref{rhoder0}.  This is convenient since it is not always possible to express $\rho$ as a function of $r$ in terms of elementary functions except for the $\mathbb{R}^2$ case. Whereas the opposite direction is straightforward. Doing this change, our metric \eqref{Gmetric} becomes
\begin{align}
ds^2=\frac{a^2}{g_0^2 \mathcal{C}^2 \mathcal{S}^2}   d\rho^2+2e^{2U(\rho)}dudv+e^{2\beta(u,\rho)}du^2 \, . \label{Gmetric2}
\end{align}
The metric function $U(r)$, using equations \eqref{U} and \eqref{rhoder0}  should satisfy
\begin{align}
\partial_\rho U(\rho) &= 2a\frac{W}{\mathcal{C}\mathcal{S}} \, , \label{UU}
\end{align}
which can be solved as
\begin{align}
U(\rho) = \begin{cases} 
      -\tfrac{1}{2}\rho^2 + c \ln \lvert \rho \rvert + 2g_0br \,, & \text{for } \mathbb{R}^2\,, \\
      -\tfrac{1}{a^2} \ln \lvert \cosh(a\rho) \rvert + c \ln \lvert \tanh(a\rho) \rvert + 2g_0br \,, & \text{for } \mathbb{H}^2, \\
       \tfrac{1}{a^2} \ln \lvert \cos(a\rho) \rvert + c \ln \lvert \tan(a\rho) \rvert + 2g_0br  \,, & \text{for } \mathbb{S}^2\,.
   \end{cases} \label{fncU}
\end{align}
where the $r$-coordinate as a function of $\rho$ is found  from \eqref{rhoder0} for $\mathbb{R}^2$   as 
\begin{align}
2g_0 r = \begin{cases} 
      \frac{1}{2\rho^2} \, ,   \text{for } \, c=0 \, , \\
       -\frac{1}{2c} \ln |\frac{c}{\rho^2}-1|  \, ,    \text{for } \, c\neq0  \, ,  \\
   \end{cases} \label{radial}
\end{align}
and for $ \mathbb{S}^2$ and  $\mathbb{H}^2$ as  \small
\begin{DispWithArrows}<2g_0 r =>[format=l]
\frac{\varepsilon a^2}{2} \ln \bigg\lvert \frac{\mathcal{C}-2x_0}{\mathcal{C}} \bigg\rvert + \varepsilon a^2x_0 \mathcal{C}^{-1}  \,, \, \text{if }  x_0=\pm \frac{1}{2a^2} \, ,  \label{radial2} \\
\frac{2\varepsilon a^2}{4a^4x_0^2-1} \ln \lvert \mathcal{C} \rvert + \frac{\varepsilon a^2}{1-2a^2x_0} \ln \lvert \mathcal{C} - x_0+ \tfrac{1}{2a^2} \rvert  + \frac{\varepsilon a^2}{1+2a^2x_0} \ln \lvert \mathcal{C} - x_0 - \tfrac{1}{2a^2} \rvert  \, ,
 \text{if }  x_0\neq \pm \frac{1}{2a^2} \, .  \notag
\end{DispWithArrows} \normalsize
The gauge field can be found from \eqref{X2} algebraically as
\begin{align}
A_u= - \frac{(\mathcal{C}-c)Q(u)}{g_0\mathcal{C}} \, , \text{where }  \partial_u \theta= Q(u) \implies 
D_u\theta=\frac{cQ(u)}{\mathcal{C}} \, .
\label{A}
\end{align}
After this, it can be shown that the gauge field equation  \eqref{feFur} is automatically satisfied using \eqref{rhoder0}. So, this solution for the vector field is only valid when $\rho$ is not equal to a constant. Also note that, for a nontrivial $A_u$, we need $c\neq 0$.  Finally, using \eqref{A} in  \eqref{Ruu0} and integrating twice we get
\begin{align}
e^{2\beta} = c_4(u) e^{2U} +c_3(u)e^{2U}\int \mathcal{C}^{-1} \mathcal{S}^{-1} e^{-2U}d\rho  -\frac{a}{g_0^2}Q^2(u) e^{2U}\int \mathcal{C}^{-3} \mathcal{S}^{-1}e^{-2U}d\rho \, ,
\end{align}
where $c_3(u)$ and $c_4(u)$ are integration constants.  Results of these integrations are too complicated to present here, but they can be seen for the $x_0=b=0$ case for the $\mathbb{S}^2$ and $\mathbb{H}^2$  sigma models in \cite{Deger:2004mw} (see its equation (3.35)), where a hypergeometric function shows up. The $c_3(u)$ and $c_4(u)$ terms can be obtained using the Garfinkle-Vachaspati method  \cite{Garfinkle:1990jq}. They correspond to waves on the background \eqref{Gmetric2} and can be removed from the solution by local coordinate transformations \cite{Gibbons:2008vi}. Meanwhile, the $Q(u)$ term corresponds to an electromagnetic wave \cite{Deger:2004mw}. Therefore, the spacetime metric \eqref{Gmetric2} describes a string superposed with waves in general. For the $\mathbb{S}^2$ and $\mathbb{H}^2$ target spaces, this string solution with $Q(u)=x_0=b=0$ appeared in \cite{Deger:1999st} and later charged version of that (i.e. $Q(u) \neq 0$)  was found in  \cite{Deger:2004mw}. 

We will now investigate the singularity structure of these string solutions.
Looking at two of their curvature invariants 
\begin{align}
&\mathcal{R}=-24g_0^2(\mathcal{C}^2+b)^2 + \frac{16g_0^2}{a^2} \mathcal{C}^2\mathcal{S}^2 \,, \notag\\
&\mathcal{R}_{\mu\nu}\mathcal{R}^{\mu\nu} =   32g_0^4 \big[ 6(\mathcal{C}^2+b)^4 + \frac{8}{a^2}(\mathcal{C}^2+b)^2
 \mathcal{C}^2\mathcal{S}^2 +\frac{3}{a^4} \mathcal{C}^4\mathcal{S}^4\big] \, ,
\end{align}
we see that they approach smoothly to the supersymmetric extremum of the potential  
\begin{align}\label{limit}
&\lim_{\rho\to \, 0}\mathcal{R}= -24g_0^2  (c^2+ b)^2\quad , &&
\lim_{\mathcal{C} \to \, 0}  \mathcal{R}= -24g_0^2 b^2  \, , \ \notag\\ 
&\lim_{\rho\to \, 0} \mathcal{R}_{\mu\nu}\mathcal{R}^{\mu\nu}= 192g_0^4 (c^2+ b)^4\quad , 
&&
\lim_{\mathcal{C} \to \, 0} \mathcal{R}_{\mu\nu}\mathcal{R}^{\mu\nu}= 192g_0^4b^4  \,.
\end{align}
The points where $g_{\rho\rho}$ component of the metric \eqref{Gmetric2}  becomes infinite are candidates for a horizon. 
It is important to note that this happens when $\mathcal{C}\mathcal{S}=0$, which is precisely where the supersymmetric extrema of the potential \eqref{potential} are located.  We will now introduce suitable coordinates for each sigma model to check this observation.  We do not need the explicit form of $e^{2\beta(u,\rho)}$ since it has no effect on curvature invariants or location of horizons. Hence, for the discussion below, we will set it to zero, which is allowed when the string has no charge.

\

\subsubsection{\texorpdfstring{$\mathbb{R}^2$}{R2} Sigma Model}


\

Here, it is convenient to work with the variable $w=\rho^{-2}\geq 0$. Now, for $c=0$ 
the string metric \eqref{Gmetric2}  becomes 
\begin{align}
ds^2= \frac{dw^2}{16g_0^2} +2 e^{-\frac{1}{w}+bw} dudv\,. \label{c=0}
\end{align}
Note that as $ w \to \infty$ ($\rho \to 0$), the metric is that of the AdS in Poincar\'e coordinates if $b\neq 0$ or Minkowski spacetime if $b=0$.  However, there is a naked singularity as $w \to 0$, which can also be seen from its curvature scalar 
\begin{align}
\mathcal{R}=- \frac{8g_0^2}{w^4}(3b^2w^4+6bw^2-8w+3) \,. 
\end{align}

On the other hand the metric for $c\neq0$ is
\begin{align}
ds^2= \frac{dw^2}{16g_0^2(1-cw)^2} +2 e^{-\frac{1}{w}} \bigg( \frac{1}{w}\bigg)^c \lvert 1-cw \rvert^{-\frac{b}{c}} dudv\,.
\end{align}
As above, it is easy to see that in the limit $ w \to \infty$ we have either AdS (when $c^2+b\neq 0$) or Minkowski spacetime (when $c^2+b=0$) and there is a singularity as $w \to 0$.  However, when $c>0$, there is now a horizon at $w=1/c$, which is precisely where the potential \eqref{potential} has another supersymmetric vacua \eqref{vacua}, namely at  $\mathcal{C}=0$ \eqref{defnC}. The near-horizon geometry is either Minkowski (for $b=0$) or AdS (for $b \neq 0$). Note that the near-horizon and asymptotic geometries can not both be Minkowski.  All these can also be seen from its scalar curvature
\small
\begin{align}
\mathcal{R}=8g_0^2\bigg[-3(b+c^2)^2+ \frac{4c(3b+2c+3c^2)}{w} - \frac{2(3b+8c+9c^2)}{w^2} + \frac{(12c+8)}{w^3} - \frac{3}{w^4} \bigg]\, .
\end{align}
\normalsize

\

\subsubsection{\texorpdfstring{$\mathbb{S}^2$}{S2} Sigma Model}


\

Recall that for $\mathbb{S}^2$, the constant $1/a^2$ is an integer.
Let us first consider the case $x_0 \neq \pm 1/(2a^2)$. We also assume $c>0$ so that the potential \eqref{potential} possesses a supersymmetric extrema at $\mathcal{C}=0$ for which we choose the $\cos(2a\rho)>0$ branch so that the constant $x_0$ is restricted as $-1/(2a^2)<x_0 <0$. After the coordinate transformation $w=\mathcal{C}^{-1}>0$ the string metric becomes
\begin{align}
ds^2 &= \frac{a^4}{g_0^2} w^{-2} \bigg[ 1-4a^4 \bigg( \frac{1}{w} -x_0 \bigg)^2 \bigg]^{-2} dw^2 \notag\\
&+ 2\bigg( \frac{1-2a^2(\frac{1}{w}-x_0)}{1+2a^2(\frac{1}{w}-x_0)} \bigg)^{x_0(1+\hat{b})} \bigg[ 1-4a^4 \bigg( \frac{1}{w} -x_0 \bigg)^2 \bigg]^{\frac{1}{2a^2}(1-\hat{b})} w^{-\frac{\hat{b}}{a^2}} dudv \label{2horizons}
\end{align}
where $\hat b=\frac{b}{(x_0^2-\tfrac{1}{4a^4})}$. Now, calculating its scalar curvature 
\begin{align}
\mathcal{R} = -\frac{8g_0^2}{a^2w^4} \big[ 3a^2b^2w^4+ (8a^4x_0^2+6a^2b-2) w^2 - 16a^4x_0w + 8a^4 +3a^2  \big] \,,
\end{align}
we determine its behavior around special points of the metric as
\begin{align}
\lim_{w\to0} \mathcal{R}= -\infty
    \,,\, \lim_{w\to\frac{2a^2}{1+2a^2x_0}} \mathcal{R}= -24g_0^2 (c^2+b)^2 \,,\, \lim_{w\to\infty} \mathcal{R}= -24g_0^2b^2 \,.
\end{align}
Clearly, there is a curvature singularity at $w=0$ and a horizon located at $w=\frac{2a^2}{1+2a^2x_0}$ (that is, $\rho=0$).
It is easy to see from the metric \eqref{2horizons} that the horizon is locally either Minkowski or AdS, depending on whether $c^2+b$ is zero or not, respectively. Asymptotic geometry ($w\to \infty$ or $\mathcal{C}\to0$) is AdS if $b\neq 0$ or Minkowski otherwise. This black string solution with $b=x_0=0$ was found in \cite{Deger:1999st} and had similar features. However, notice that in that case the asymptotic limit is necessarily Minkowski, and the near-horizon geometry must be AdS. Finally, let us note that when $b=x_0=0$ and $a^2=1/2$ the solution \eqref{2horizons} becomes identical with the one found in \cite{Horne:1991gn} for the low energy limit of the
3D String theory as noted in \cite{Deger:1999st}.

When $x_0=1/(2a^2)$ we have $c=1/a^2$ and $\mathcal{C}=\cos^2(a\rho)/a^2$. After the coordinate transformation $w=\sec^2(a\rho)$ string metric \eqref{Gmetric2} becomes
\begin{align}
ds^2 = \frac{a^4}{16g_0^2} \bigg( 1-\frac{1}{w} \bigg)^{-2} dw^2 + 2\bigg( 1-\frac{1}{w} \bigg)^{\frac{1}{a^2}} (w-1)^{a^2b} e^{a^2bw} dudv \, ,
\end{align}
whose scalar curvature is
\begin{align}
\mathcal{R} = -\frac{8g_0^2}{a^8w^4} \big[ 3a^8b^2w^4+ 6a^4b w^2 - 8a^2w + 8a^2 +3  \big] \, .
\end{align}
It behaves as
\begin{align}
&\lim_{w\to0} \mathcal{R}= -\infty
  \,,&&\lim_{w\to1} \mathcal{R}= -24g_0^2(c^2+b)^2  \,,
&& \lim_{w\to\infty} \mathcal{R}= -24g_0^2b^2 \, .
\end{align}
Similar to the above solution, there is a curvature singularity at $w=0$, a horizon at $w=1$ (or $\rho=0$), and the asymptotic geometry ($w\to \infty$) is either Minkowski or AdS. 

When $x_0=-1/(2a^2)$, we have $c=0$ and hence there is only one supersymmetric vacuum of the potential \eqref{potential}. Thus, the corresponding string solution has a naked singularity. 

\subsubsection{\texorpdfstring{$\mathbb{H}^2$}{H2} Sigma Model}

Let us first assume $x_0 \neq \pm 1/(2a^2)$ and take $x_0>1/(2a^2)$. Defining $w=\sinh^{-2}(a\rho)$ its metric becomes
\begin{align}
ds^2 &= \frac{1}{16g_0^2} \bigg( 1+\frac{1}{w} \bigg)^{-2} w^{-2} \bigg[ -\frac{1}{2a^2}\bigg(1+\frac{2}{w} \bigg) +x_0 \bigg]^{-2} dw^2 \\
& + 2 ( w+1)^{-x_0(1+\hat{b})} \bigg(\frac{4 (w+1)}{w^2}\bigg)^{-\frac{1}{2a^2}(1-\hat{b})} \bigg[ -\frac{1}{2a^2}\bigg(1+\frac{2}{w} \bigg) +x_0 \bigg]^{-\frac{\hat{b}}{a^2}} dudv \, , \notag
\end{align}
where $\hat b=\frac{b}{(x_0^2-\tfrac{1}{4a^4})}$. Its scalar curvature is too long to present here, but it has the limits
\begin{align}
& \lim_{w\to0}  \mathcal{R}= \pm\infty
 \,,&& \lim_{w\to\frac{2}{2a^2x_0-1}}  \mathcal{R}= -24g_0^2b^2 \,,&&  \lim_{w\to\infty} \mathcal{R}= -24g_0^2 (c^2 +b)^2 \,.
\end{align}
There is a curvature singularity at $w=0$ (carrying the sign of $(8a^2-3)$), a horizon at $w=\frac{2}{2a^2x_0-1}$ (or $\mathcal{C}=0$) and asymptotic geometry ($w\to \infty$ or $\rho \to 0$) is Minkowski or AdS. This solution with $ x_0 = b = 0$ was obtained in \cite{Deger:1999st}, but in that case there is no horizon and the asymptotic geometry is AdS.

For $x=\pm 1/(2a^2)$, the potential \eqref{potential} has only one supersymmetric extremum at $\rho=0$ and hence the string solution has a naked singularity.

\section{Timelike Killing Vector} \label{sec5}
We will now investigate the timelike case and take $f>0$. The most general spacetime metric that admits a timelike Killing vector  $V=\partial_t$ with norm $-4f^2$ can be written as a product of a 2-dimensional base space  $\Sigma_2$ and a timelike fiber \cite{Deger:2013yla}
\begin{align}
ds^2=-4f^2\bigl[dt+M(x,y)dx+N(x,y)dy\bigr]^2+e^{2\sigma(x,y)}(dx^2+dy^2) \,. \label{timelikemetric}
\end{align}
The base space $\Sigma_2$ has constant Gaussian curvature $\kappa$ if the conformal factor $\sigma$ satisfies Liouville's differential equation
\begin{align}
(\partial_x^2 +\partial_y^2)\sigma= -\kappa e^{2\sigma} \, . \label{Liouville} \
\end{align}
With this metric, the conditions  \eqref{algebraic} can be solved as
\begin{align}
K_\mu=(0,2fe^\sigma\sin\varphi,2fe^\sigma\cos\varphi)
 \, , \quad L_\mu=(0,2fe^\sigma\cos\varphi,-2fe^\sigma\sin\varphi) \, ,
\end{align}
where $\varphi(t,x,y)$ is an arbitrary function to be determined. Then the Killing spinors are found to be \cite{Deger:2024obg}
\begin{align} \label{Killing3}
   \lambda_1=(1+i)\, \sqrt{\frac{f}{2}} \begin{bmatrix} \sin(\varphi/2) \\ -\cos(\varphi/2) \end{bmatrix} \, , \quad \lambda_2=(1+i)\, \sqrt{\frac{f}{2}} \begin{bmatrix} \cos(\varphi/2) \\ \sin(\varphi/2) \end{bmatrix} \, .
\end{align}
From these Majorana spinors, one gets the complex Killing spinor of these solutions \eqref{sp} as 
\begin{align} \label{Killing4}
   \epsilon=(1+i)\, \sqrt{\frac{f}{2}} e^{-i\varphi/2} \begin{bmatrix} i \\ -1 \end{bmatrix} \,  .
\end{align}
Equations \eqref{kd} and \eqref{ld} determine $X_\mu$ defined in \eqref{X}  as
\begin{align}
& X_t \,=\, \tfrac{1}{2}\partial_t\varphi + 4fg_0W \,, \notag\\
 &X_x \,=\, \tfrac{1}{2}\partial_x\varphi +\frac{1}{2}\partial_y\sigma + 4fg_0MW \,, \notag\\
 &X_y \,=\, \tfrac{1}{2}\partial_y\varphi -\frac{1}{2}\partial_x\sigma + 4fg_0NW \,.  \label{Xeqns}
\end{align}
Note that because of \eqref{Lie1} and \eqref{Lie2}, scalar fields and the vector field do not depend on the time coordinate $t$. Our gauge choice \eqref{gauge} implies that
\begin{align}
A_t =4f\mathcal{C}\,. \label{Atcomp}
\end{align}
Using this result and \eqref{X} in \eqref{Xeqns} we get
\begin{align}
\partial_t \varphi &= -8fg_0b \,, \label{newXeqns1}\\
D_x\theta \, \mathcal{C}   &\,=\, \partial_x(c\theta+\frac{1}{2}\varphi) +\frac{1}{2}\partial_y\sigma + 4fg_0MW \,, \label{newXeqns2}\\
D_y\theta \, \mathcal{C}   &\,=\, \partial_y(c\theta+\frac{1}{2}\varphi) - \frac{1}{2}\partial_x\sigma + 4fg_0NW \,. \label{newXeqns3}
\end{align} 
From \eqref{vd} we derive
\begin{align}
W\,=\,-\frac{1}{2g_0}f e^{-2\sigma}(M_y-N_x) \label{Weqn} \, ,
\end{align}
This completes the information available in the Killing spinor equation \eqref{kse}.
Next, from the BPS conditions \eqref{bps2}-\eqref{bps3} we find
\begin{align}
&\partial_x\rho= -\frac{1}{2a} \mathcal{S}(D_y\theta- 4fg_0\mathcal{C}N) \,,  \notag \\
&\partial_y\rho=  \frac{1}{2a} \mathcal{S}(D_x\theta - 4fg_0\mathcal{C}M) \,. \label{timelikeBPS}
\end{align}
Now, one can show that when the above first-order conditions are satisfied, all field equations \eqref{fieldeq} are also satisfied except the $xy$-component of the vector field equation
\begin{align}
F_{xy}= \frac{2}{a^2} \mathcal{S}^2 \big\{ f\big[ N D_y\theta+M D_x\theta -4fg_0\mathcal{C}(M^2+N^2) \big] + g_0\mathcal{C}e^{2\sigma} \big\} \,. \label{FFeqnEquiv}
\end{align}
To summarize, for a supersymmetric solution in this class, we have to solve 7 first order, coupled differential equations  \eqref{newXeqns1}-\eqref{FFeqnEquiv} for 8 unknowns $(M, N, \sigma, \varphi, A_x, A_y, \theta, \rho)$. Although, this system looks like under determined at first sight, but after defining a new spinor function  $\hat{\varphi}=c\theta+\frac{1}{2}\varphi$ (recall that $\partial_t \theta=0$) and using the fact that $g_0F_{xy}=\partial_x(D_y\theta) -\partial_y(D_x\theta)$, we end up with 7 differential equations for 7 unknowns  $(M, N, \sigma, \hat{\varphi}, D_x\theta, D_y\theta, \rho)$.
Solving this system in full generality is clearly a difficult problem except when $\rho$ is equal to a constant. This is what we will consider in the next two subsections, and then for the general case, make a simplifying ansatz.

\subsection{\texorpdfstring{$\rho\,=0$}{rho=0}}
When $\rho=0$, we have conditions listed in \eqref{rho=0}.
The field $\theta$ disappears from all the equations \eqref{newXeqns1}-\eqref{FFeqnEquiv} and therefore remains free.  Moreover, equations in \eqref{timelikeBPS} are automatically satisfied, and \eqref{FFeqnEquiv} gives $F_{xy}=0$. So, we can choose $A_x=A_y=0$ without loss of generality. 
The only equations to be solved are \eqref{newXeqns1}-\eqref{newXeqns3} which simplify to
\begin{align}
0 &= \partial_t \varphi +8fg_0b \,, \notag\\ 
0 &= \partial_x \varphi  +\partial_y \sigma  +8fg_0WM  \,, \notag\\ 
0 &= \partial_y \varphi  -\partial_x \sigma +8fg_0WN \,. \label{rho0XEqns}
\end{align}
Let us point out that combining the last two and using \eqref{Weqn}, one gets Liouville's equation \eqref{Liouville} for $\sigma$ with Gaussian curvature $\kappa= -16g_0^2W^2$, which shows that the base space $\Sigma_2$ of the metric \eqref{timelikemetric}  is either flat or hyperbolic, depending on whether $W$ is vanishing or not. We now analyze these two possibilities separately.

\newpage

\textbf{\underline{Case 1: $W\neq0$}}

\

In this case, we have $W=c^2+b\neq 0$.
The first equation in \eqref{rho0XEqns} gives
\begin{align}
\varphi = -8fg_0b\,t + G(x,y) \, ,
\end{align}
for some function $G(x,y)$. But, using a new time coordinate $\hat{t} = t - \frac{1}{8fg_0W}G(x,y)$,  and redefining metric functions $M$ and $N$ we can set $G(x,y)=0 \implies \partial_x \varphi=\partial_y \varphi=0 $, as in \cite{Deger:2024obg}. Remaining equations can be solved as
\begin{align}
\sigma &= -\ln \big\lvert 1- 4g_0^2W^2 (x^2+y^2) \big\rvert \, , \notag \\
M &= -\frac{g_0 W}{f} \frac{y}{1- 4g_0^2W^2 (x^2+y^2)}  \,, \notag\\
N &= \frac{g_0 W}{f}  \frac{x}{1- 4g_0^2W^2 (x^2+y^2)}   \,, \label{globalAdS}
\end{align}
and the scalar curvature is $\mathcal{R}= -24 g_0^2 (c^2+b)^2$.  The solution is $AdS_3$ in global coordinates, as was shown in \cite{Deger:2024obg}.

\

\textbf{\underline{Case 2: $W=0$}}

\

This corresponds to setting $W=c^2+b= 0$, which also makes the potential \eqref{potential} vanish, i.e. $V=0$. Since all scalar and vector fields also vanish, these are pure gravity solutions and are Ricci flat. In this case, \eqref{rho0XEqns} become
\begin{align}
&\partial_t\varphi = 8fg_0c^2 \,, \notag \\
&\partial_x \varphi = -\partial_y \sigma \,, \notag\\
&\partial_y \varphi = \partial_x \sigma \, .
\end{align}
The first one gives
\begin{align}
\varphi = 8fg_0c^2\,t +G(x,y)\, . \label{spin}
\end{align}
Note that both $G$ and $\sigma$ are harmonic functions and they can be used to define a  holomorphic function 
\begin{align}
H(z) = \sigma + iG \, , \label{H}
\end{align}
where $z=x+iy$. From \eqref{Weqn}, we see that $Mdx+Ndy$ is closed. If we assume that the space part of our solution is simply connected, then the Poincar\'e lemma implies that we can find a function $F(x,y)$ such that $Mdx+Ndy=dF$. But then, we can define a new time coordinate $\hat{t} = 2ft +F(x,y)$ to get rid of $F(x,y)$ and redefine $G(x,y)$. After doing this, the metric takes the form
\begin{align}
ds^2= -dt^2 + e^{2\sigma}(dx^2+dy^2) \,. 
\end{align}
Any choice of the holomorphic function $H(z)$ gives a supersymmetric solution with the above metric, as happened in  \cite{Deger:2024obg} (see its equations (6.28)-(6.30)).  For example,the choice $H(z)=x+iy$ leads to
\begin{align}
ds^2= -dt^2 + dw^2+w^2dy^2 \,, \label{ERspace}
\end{align}
where $w=e^x$. This solution can be designated as Euclidean Rindler spacetime \cite{Deger:2006uc}. If the range of $y$ is restricted to $0\leq |y| < 2\pi$, then it describes a conical spacetime.

\subsection{\texorpdfstring{$\rho=constant\neq0$}{rho const}}
In this case, from \eqref{timelikeBPS} we can read the gauge field components as 
\begin{align}
& A_x = -\tfrac{1}{g_0}\partial_x\theta +4f\mathcal{C}M \,,   \notag\\
& A_y = -\tfrac{1}{g_0}\partial_y\theta +4f\mathcal{C}N \,.  \label{gfc}
\end{align}
The remaining equations to be solved are 
\begin{align}
&\partial_t \hat\varphi = -8fg_0b \,, \notag\\
&\partial_x \hat\varphi + \partial_y \sigma = -8fg_0bM \,, \notag\\ 
&\partial_y \hat\varphi - \partial_x \sigma = -8fg_0bN  \,, \label{weirdCaseeqns}
\end{align}
where $\hat\varphi = 2c\theta + \varphi$.
The last two lines give Liouville's equation \eqref{Liouville} for $\sigma$ with Gaussian curvature $\kappa= -16g_0^2bW$.
In all the solutions we present below, we have $\kappa \leq 0$, that is, $Wb \geq 0$.

Since $\rho \neq 0$ we have  $\mathcal{S}\neq 0$ and using \eqref{timelikeBPS} in \eqref{FFeqnEquiv} one finds
\begin{align}
\mathcal{C} W= \frac{\mathcal{C}\mathcal{S}^2}{4a^2} \, ,
\end{align}
which implies that either $\mathcal{C}=0$ or $W=\mathcal{S}^2/4a^2$. For the latter,  the potential \eqref{potential} has the value $V=2W(2b-W)$ and this point is not an extremum in general, but this happens when $\mathcal{C}=x_0$ for the $\mathbb{S}^2$ manifold, which occurs at $\rho=\pm \pi/(4a)$ \eqref{defnC}. This condition, together with the definition of the superpotential \eqref{superpotential}, fixes the value of $\rho\neq 0$ unless $x_0=0, a^2=-4b=-\varepsilon=1$, for which it is automatically satisfied. 

\bigskip

\textbf{\underline{Case 1: $\mathcal{C}=0$}}

\

When  $\mathcal{C}=0$ and $\rho \neq 0$, from \eqref{defnC} we see that $c\neq 0$ and $x_0\neq 0$ \eqref{x0}
and $\mathcal{S}$ has the same value \eqref{sigma}. Note that 
since  the superpotential \eqref{superpotential} is  now $W=b$, the base space $\Sigma_2$ of our metric \eqref{timelikemetric} has Gaussian curvature $\kappa=-16g_0^2b^2$. So, it is either flat or hyperbolic. This case is very similar to the $\rho=0$ case that we considered above.  For $b\neq 0$, making the change $W\rightarrow b$ in the AdS solution \eqref{globalAdS}  is enough. For $b=0$, instead of \eqref{spin}, we now have $\hat\varphi= G(x,y)$, and the rest is the same. These solutions emerge at the supersymmetric extrema of the potential \eqref{potential}.

\

\textbf{\underline{Case 2: Timelike Warped Flat}}

\

We now assume $W=\mathcal{S}^2/(4a^2)\neq 0$, $\mathcal{C} \neq 0$ and $b=0$.
From \eqref{weirdCaseeqns} we see that functions $\hat\varphi$ and $\sigma$, define a holomorphic function
\begin{align}
H(z)= \sigma + i \hat\varphi \, .
\end{align}
Any choice of $H(z)$ gives a supersymmetric solution by solving the first line of \eqref{weirdCaseeqns} for $M$ and $N$.
Since $\sigma$ is a harmonic function, base space $\Sigma_2$ of our metric \eqref{timelikemetric} is always flat from \eqref{Liouville} for any choice of the function $H(z)$ and spacetime curvature is always positive:
\begin{align}
\mathcal{R}= 8g_0^2W^2 \, .
\end{align}
For example, by choosing $H(z)=0$ and setting $\theta=N=0$ the solution is
\begin{align} \label{twflat}
ds^2 = - (2fdt - \frac{2g_0W}{f} ydx)^2 + dx^2+dy^2 \quad , \quad A_\mu = (4f\mathcal{C}, -\frac{8g_0W}{f} \mathcal{C}  y, 0) \, ,
\end{align}
which is called timelike warped flat \cite{Deger:2016vrn, Chow:2019ucq} due to the Hopf fibration, and it is a homogeneous spacetime \cite{Charyyev:2017uuu, Corral:2024xfv}. Since $W=\mathcal{C}^2\neq 0$, there is always warping, and it is proportional to $A_\mu A^\mu=-4 \mathcal{C}^2$.

\

\textbf{\underline{Case 3: Timelike stretched AdS$_3$}} 

\

We now assume  $W=\mathcal{S}^2/(4a^2)\neq 0$,  $\mathcal{C}\neq 0$ and $b\neq0$.  In solving the system \eqref{weirdCaseeqns}, we can again set 
\begin{align}
\theta=0 \quad , \quad \hat{\varphi}=\varphi= -8g_0bt \, ,
\end{align}
by shifting the time coordinate and making a gauge transformation. Then, one can start solving from  Liouville's equation  \eqref{Liouville}  for $\sigma$ with $\kappa=-16g_0^2bW$. For any $\sigma$, the spacetime curvature is always constant:
\begin{align}
\mathcal{R} = 8 g_0^2W(W-4b)\,.
\end{align}
For example, equations \eqref{weirdCaseeqns} can be solved as
\begin{align}
e^{2\sigma} = \frac{1}{16g_0^2bW}\frac{1}{y^2} \quad , \quad 
 M = \frac{1}{8fg_0b} \frac{1}{y}\quad , \quad N=0 \,,
\end{align}
with the requirement
\begin{align}
Wb >0\, ,
\end{align}
which implies that the base space $\Sigma_2$ is hyperbolic.
Since $W=\mathcal{S}^2/4a^2$ is positive, we need $b>0$. The metric is (after scaling the time coordinate as $\hat{t}=8fg_0b t$ and then dropping  the hat)
\begin{align}\label{twAdS}
ds^2= \frac{1}{16g_0^2bW} \bigg[ - \nu^2 \bigg(dt + \frac{dx}{y}\bigg)^2+\frac{dx^2+dy^2}{y^2}\bigg] \,,
\end{align}
where
\begin{align}
 \nu^2= \frac{W}{b}= 1+ \frac{\mathcal{C}^2}{b}  \,.
\end{align}
This is the "stretched"  timelike warped AdS$_3$ with the warping parameter $\nu^2$.  The term "stretched" refers to the fact that   $\nu^2>1$ (since $b>0$ and $\mathcal{C}\neq 0$). The gauge field is given as 
\begin{align}
A_\mu=(\frac{\mathcal{C}}{2g_0b}, \frac{\mathcal{C}}{2g_0b\,y}, 0) \, .
\end{align}
One can relate the warping parameter to the norm of the gauge field $A^2= A_\mu A^\mu$  as
\begin{align}
\nu^2= 1- \frac{A^2}{4b} \, ,
\end{align}
which was first observed in \cite{Deger:2013yla} where this supersymmetric solution was found for an off-shell supergravity (see also \cite{Deger:2016vrn, Deger:2018kur}).   Note that, since $\mathcal{C}\neq 0$ and $\rho\neq 0$, this solution does not appear at a supersymmetric extrema of the potential \eqref{potential}. 

\bigskip

\subsection{Circularly Symmetric Solutions}

We now assume that $\rho$ is not constant. To simplify the equation system \eqref{newXeqns1}-\eqref{FFeqnEquiv}, we will take the $x$-coordinate as a circular direction and consider the circularly symmetric ansatz:
\begin{align}
&\rho=\rho(y) \,, &&\sigma=\sigma(y) \,, &&M=M(y) \,, &&N=A_y=0 \,, \notag\\
&A_t =4f\mathcal{C} \,, && A_x=\chi(y)+\tfrac{n}{g_0} \,, &&\varphi = -4fg_0bt + kx \,,  &&\theta=-nx \,, 
\end{align}
where $n$ and $k$ are constants. Then we are left with the following system of equations
\begin{align}
\partial_y\rho &= \frac{g_0}{2a} \mathcal{S} \tilde{\chi} \,, \label{c1}\\
\partial_y M &= -\frac{2g_0}{f}e^{2\sigma}W\,, \label{c2}\\
\partial_y \sigma &= 2g_0( \mathcal{C}\tilde{\chi} -4fbM) +2cn -k \,, \label{c3}\\
\partial_y \tilde{\chi} &= 2g_0 \mathcal{C} e^{2\sigma}  \left(4W -\frac{\mathcal{S}^2}{a^2} \right) \, ,
\label{c4}
\end{align}
where we defined
\begin{align} \label{chi}
\tilde{\chi}= \chi-4f\mathcal{C}M \, .
\end{align}
These equations were first derived in \cite{Abou-Zeid:2001inc} and later studied in \cite{Deger:2006uc} for the $x_0=b=0$ case.
 In \cite{Abou-Zeid:2001inc} it was shown that this system admits $|n|$-vortex solutions
using numerical methods. In these vortices, the function $\chi$ is not constant and solutions have asymptotically conical geometry for $b=0$ and $2cn-k \neq 0$.  By differentiating \eqref{c2} and using others, one arrives at the following constraint for the unknown functions
\begin{align} \label{p}
\tilde{\chi}^2 + 16f^2bM^2+  \frac{4f}{g_0} (k-2cn)M -4e^{2\sigma}W =p \, , 
\end{align}
where $p$ is a constant, as was observed in \cite{Abou-Zeid:2001inc}. This condition reduces the number of unknowns to 3.  Furthermore, one can use $\rho$ as the independent variable instead of $y$, using \eqref{c1}. Then equations \eqref{c2}-\eqref{c4} take the form
 \begin{align}
 \mathcal{S} \tilde{\chi} \partial_\rho M  &= -\frac{4a}{f}e^{2\sigma}W\,, \label{c22}\\
 \mathcal{S} \tilde{\chi} \partial_\rho \sigma &= 4a( \mathcal{C}\tilde{\chi} -4fbM) +2cn -k  \,, \label{c33}\\
 \mathcal{S} \tilde{\chi} \partial_\rho \tilde{\chi} &= 4a \mathcal{C} e^{2\sigma}  \left(4W -\frac{\mathcal{S}^2}{a^2} \right) \, ,
\label{c44}
\end{align}
So, effectively, there are only 2 first-order differential equations to solve, but due to their nonlinear structure, this is still a difficult problem. Below we will find exact solutions when the vector field function $\chi$ is a constant. 
Let us point out that there are two other special choices that simplify the equation system \eqref{c22}-\eqref{c44}. When $b=0$ equation \eqref{c33} and when $a^2=-4b=-\varepsilon=1$ ($\implies 4Wa^2=\mathcal{S}^2)$ equation \eqref{c44} become integrable. However, in the end, these lead to the solution with $\chi=0$ below, with the parameters specified to those particular values.

\

\textbf{\underline{Case 1: String-like Solutions}}

\

We now assume $\chi=0$, that is,
\begin{align} \label{chi2}
\chi=0 \implies  \tilde{\chi}= -4f\mathcal{C}M \, .
\end{align}
Equation \eqref{c4} is equivalent to
\begin{align}
\partial_y \chi &= -\frac{2g_0}{a^2} \mathcal{S}^2  \big\{ fM(\chi-4f\mathcal{C}M) + \mathcal{C}e^{2\sigma} \big\} 
 \label{c5} \, .
\end{align}
So, when $\chi=0$ we have
\begin{align}
    e^{2\sigma}=4f^2M^2 \, . \label{sigma2}
\end{align} 
With \eqref{chi2} and \eqref{sigma2}, equation \eqref{c44} is automatically satisfied, and the constraint \eqref{p} holds with $p=0$. Then, compatibility of \eqref{c22} and \eqref{c33} requires 
\begin{align}
2cn -k =0 \, .
\end{align}
So, we only have to solve \eqref{c22} or \eqref{c33}. But comparing \eqref{c33} with \eqref{UU} we see that
\begin{align}
\sigma=2U \, .
\end{align}
Then, the solution metric becomes (after some trivial scalings)
\begin{align} \label{metrictimelike}
ds^2 = - dt^2 - 2 e^{2U} dxdt + \frac{a^2}{g_0^2} \frac{d\rho^2}{\mathcal{S}^2\mathcal{C}^2} \,,
\end{align}
where the function $U$ is given in \eqref{fncU}, \eqref{radial} and \eqref{radial2}. For the $\mathbb{S}^2$ and $\mathbb{H}^2$ sigma models, this solution with $x_0=b=0$ was found in \cite{Deger:2006uc}. Notice that the form of the metric is almost the same with the string solution \eqref{Gmetric2}, and due to this similarity, we will call this solution "string-like" as in \cite{Deger:2006uc}. The extra $-dt^2$ term in this metric has no effect on curvature invariants like the $e^{2\beta}$ piece there, and therefore, these two solutions have the same singularity and horizon structure that we studied in section \ref{sectstr}.  
However, when the local geometry is AdS, due to this extra piece, the metric is of Kaigorodov type \cite{Cvetic:1998jf, Brecher:2000pa} (see \cite{Deger:2004mw, Deger:2006uc}). 

\newpage

\textbf{\underline{Case 2: $\chi=constant \neq 0$}}

\

Let $\chi=\chi_0 \neq 0$ be a constant. Again, setting $2cn-k=0$ is necessary. In this case \eqref{c5} gives
\begin{align}
e^{2\sigma} = -\frac{ fM \tilde{\chi}}{ \mathcal{C}} \, . 
\end{align}
Now, one can compute $\partial_\rho M$ from both \eqref{c33} and \eqref{c44} and they are compatible only for the $\mathbb{H}^2$ manifold with the following choices: 
\begin{align}
a^2=4b=-\varepsilon= 1 \, . \label{para}
\end{align}
Assuming $x_0=0$, the corresponding solution is
\begin{align}
e^{2\sigma}= \chi_0^2 \mathcal{S}^{-4} \, , \,  \tilde{\chi} = -\chi_0 \mathcal{S}^{-2}  \, , \, M=  \frac{\chi_0} {f} \mathcal{S}^{-2} \mathcal{C} \, .
\end{align}
The constraint \eqref{p} is satisfied with $p= 0$ as above. Note that it is not possible to take $\chi_0 \rightarrow 0$ limit. 
Although changing $\chi$ from zero to a nonzero value $\chi_0$ is a trivial gauge transformation, we see that it results in a solution that is quite different than the $\chi=0$ solution with the same parameters \eqref{para} for which we have  $e^{2\sigma}= 4M_0^2 \mathcal{S}^{-4} \mathcal{C}^2, \,  \tilde{\chi} = -4M_0 \mathcal{S}^{-2} \mathcal{C}^2, \,   M=M_0\mathcal{S}^{-2} \mathcal{C}/f$. This is due to the nonlinear nature of these differential equations.  
After defining $g_0w=\ln \lvert \tanh(a\rho) \rvert$ and some straightforward scalings, its metric becomes
\begin{align} \label{atimelikesoln}
ds^2 = -dt^2  + 2 \cosh(2g_0w) dxdt -4 \sinh^2 (g_0w) dx^2 + dw^2 \,,
\end{align}
with scalar curvature
\begin{align}
\mathcal{R} = \frac{g_0^2}{2} \frac{\cosh (2g_0w)}{\sinh^4 (g_0w)} (4-3\cosh(2g_0w)) \, .
\end{align}
When $w\rightarrow \infty$ (or $\rho \to 0$) its curvature scalar takes the value $\mathcal{R}=-6g_0^2$
and there is a naked singularity as $w\rightarrow 0$ (or $\rho \rightarrow \infty$).  Unfortunately, we could not identify this spacetime.

\section{Conclusions} \label{sec6}

In this paper, we studied supersymmetric solutions of the 3D $N=2$ gauged supergravity extended with FI-term, which was constructed in \cite{Abou-Zeid:2001inc}. Using Killing spinor bilinears, we first derived all the necessary conditions for such a solution to exist and then constructed many examples, which are listed in Table \ref{table}. We expect them to be useful in various contexts. Among them, we have supersymmetric examples of timelike \eqref{twAdS} and null z-warped AdS \eqref{nwAdS}
which are quite rare in the literature. They require FI-term to be present, and curiously, only the latter appears at a supersymmetric extremum. As we showed, parameters of the model allow the null z-warped AdS solution to be converted into a well-defined black hole, and understanding holographic aspects of it is clearly desirable \cite{Anninos:2010pm}. The fact that the scaling exponent $z$ is free provides an opportunity to go beyond the better understood $z=2$ case \cite{Guica:2010sw}.

Another interesting outcome of our work is the string \eqref{2horizons} and string-like \eqref{metrictimelike}
solutions with horizons located at the supersymmetric vacua of the potential which has AdS or Minkowski geometry. To clarify their global structure and conserved charges \cite{Hajian:2016iyp}, more work is needed. A detailed analysis of geodesics \cite{Horne:1991gn} would help in understanding whether their singularity is physically acceptable or not \cite{Gubser:2000nd}. Moreover, our exact black string solutions can also be used to analyze holographic renormalization group flows as in \cite{Berg:2001ty, Deger:2002hv, Gava:2010rx,
Arkhipova:2024iem, Golubtsova:2024dad, Gutperle:2024yiz}. Those that connect two AdS vacua are especially interesting to explore.

A natural continuation of the present work is to construct supersymmetric solutions of the ungauged version of this model. 
This problem was studied for the ungauged limit of a 3D, $N=4$ model in \cite{Deger:2024obg} for which the target space of the sigma model is $\mathbb{H}^2$. It was found that for timelike solutions, the sigma model target space metric is identified with the space part of the spacetime metric, and solutions are characterized in terms of two holomorphic functions. The model we considered in this paper provides an opportunity to explore this phenomenon for other sigma model geometries.

Embedding this model into 3D gauged supergravities with more supercharges also remains to be done. Finding its higher-dimensional origin looks even more challenging \cite{OColgain:2010yxb, Karndumri:2013dca, OColgain:2015dbh,  Karndumri:2015sia},
especially when the sigma model target space is compact.
We hope to examine these issues in the near future.

\section*{Acknowledgments}
It is a pleasure to thank H. Samtleben and M.M. Sheikh-Jabbari for useful correspondence. We would like to thank ENS de Lyon for hospitality during the course of this work.
NSD is grateful to Albert Einstein Institute, Potsdam, and Rudjer Boskovic Institute, Zagreb, for hospitality where part of this work was carried out. Both authors are partially supported by 
the Scientific and Technological Research Council of T\"urkiye (T\"ubitak) project 123N953.


\bibliographystyle{utphys} 
\bibliography{references.bib}

\end{document}